\documentclass[12pt, letterpaper]{extarticle}

\usepackage{subcaption}
\usepackage{fullpage}
\usepackage[switch]{lineno}
\usepackage{amsmath}
\usepackage{amssymb}
\usepackage{rotating}
\usepackage{array}
\usepackage{mathtools}
\usepackage[ruled]{algorithm2e}
\usepackage{algorithmic}
\usepackage{bm}
\usepackage{breqn}
\usepackage{comment}
\usepackage{enumitem}
\usepackage{graphics}
\usepackage{graphicx}
\usepackage{latexsym}
\usepackage{mathrsfs}
\usepackage{morefloats}
\usepackage{nicefrac}
\usepackage{authblk}
\usepackage{pifont}

\usepackage[hyphens]{url}
\usepackage{hyperref}
\hypersetup{colorlinks=false,breaklinks=true}

\title{Inclusive content reduces racial and gender biases, yet non-inclusive content dominates popular culture}

\author[1+]{Nouar AlDahoul}
\author[1+]{Hazem Ibrahim}
\author[1*]{Minsu Park}
\author[1*]{Talal Rahwan}
\author[1*]{Yasir Zaki}

\affil[1]{\normalsize New York University Abu Dhabi, UAE.}
\affil[+]{\footnotesize Joint first authors ordered alphabetically.}
\affil[*]{\footnotesize Corresponding authors. E-mail: \{minsu.park,talal.rahwan,yasir.zaki\}@nyu.edu}
\date{}

\linenumbers

\begin{document} 
\maketitle 
\vspace{0.5cm}
\begin{abstract}
\noindent 
Images are often termed as representations of perceived reality. As such, racial and gender biases in popular culture and visual media could play a critical role in shaping people's perceptions of society. While previous research has made significant progress in exploring the frequency and discrepancies in racial and gender group appearances in visual media, it has largely overlooked important nuances in how these groups are portrayed, as it lacked the ability to systematically capture such complexities at scale over time. To address this gap, we examine two media forms of varying target audiences, namely fashion magazines and movie posters. Accordingly, we collect a large dataset comprising over 300,000 images spanning over five decades and utilize state-of-the-art machine learning models to classify not only race and gender but also the posture, expressed emotional state, and body composition of individuals featured in each image. We find that racial minorities appear far less frequently than their White counterparts, and when they do appear, they are portrayed less prominently. We also find that women are more likely to be portrayed with their full bodies, whereas men are more frequently presented with their faces. Finally, through a series of survey experiments, we find evidence that exposure to inclusive content can help reduce biases in perceptions of minorities, while racially and gender-homogenized content may reinforce and amplify such biases. Taken together, our findings highlight that racial and gender biases in visual media remain pervasive, potentially exacerbating existing stereotypes and inequalities.
\end{abstract}

\section*{Introduction}
The significance of racial and gender representation in popular culture and visual media extends far beyond a simple reflection of demographic realities. Media play a critical role in shaping societal norms and perceptions, as they have the power to either challenge or reinforce stereotypes and biases deeply ingrained in both majority and minority groups. For example, the diversity in the depiction of race and gender can challenge harmful stereotypes, fostering a broader recognition of social and cultural plurality~\cite{mastro2015media, tukachinsky2015we, browne1999television, ramasubramanian2011impact, shinoda2021beyond, ward2020media}. Positive and nuanced portrayals of individuals across different racial and gender identities can reduce prejudice, encourage empathy, and disrupt long-standing societal biases~\cite{browne1999television, clark2019seeing}. 

Over the past several decades, a growing body of literature has aimed to quantify the representation of race and gender in popular culture and visual media. Much of this research has examined how racial minorities and different genders are portrayed, often focusing on patterns of body composition~\cite{mitchell2023representation, millard2006stereotypes, lindner2004images}, representation~\cite{topaz2022race,reddy2018critical, seiter1990different, smith2012gender, smith2010assessing}, and their impact on public perceptions of reality~\cite{eschholz2002symbolic, berger2016social,wardle1999representation,cavender2018television,stevenson2002understanding, gitlin1986watching}. However, these studies have been constrained by narrow temporal scopes, typically spanning only a few years, and by a focus on prominent formats such as magazine covers or high-profile films. This narrow scope often overlooks smaller in-page images or less prominent visual media that could provide a more comprehensive understanding of racial and gender disparities across a wider array of content. Furthermore, the labor-intensive nature of manually coding socio-demographic identities, poses, and other visual attributes has limited the scale of many studies. As a result, much of the literature lacks the large-scale, longitudinal analysis necessary to capture how these portrayals evolve over time.

In this study, we conduct a large-scale longitudinal analysis of racial and gender bias in two distinct yet pervasive visual media formats: fashion magazines and movie posters. We select these media forms because they target different audience segments---fashion magazines typically cater to niche, trend- and beauty-conscious groups, while movie posters reach a much broader and more general audience. Both are highly visual and widely consumed, making them well-suited for studying how racial and gender biases are reflected and reinforced in everyday life. Furthermore, fashion magazines and movie posters offer rich, curated imagery that is specifically designed to project cultural standards~\cite{law2002cultural, rhodes2024film}, providing a unique lens for investigating the role of stereotypical portrayals of race and gender. 

We analyze a dataset comprising over 300,000 images spanning five decades and employ state-of-the-art machine learning classifiers to systematically detect not only race and gender but also posture, body composition, and the emotional states expressed by individuals depicted in visual media (see Methods for more details). By leveraging these datasets and techniques, we aim to address existing gaps and conduct more comprehensive studies across longer time-frames and a broader range of media formats. Additionally, we validate our findings using a third independent, well-known public dataset of advertisements~\cite{hussain2017automatic}. While this dataset may be less representative, it still can offer important insights, as advertisements are ubiquitous and play a key role in shaping idealized representations that directly appeal to consumer desires~\cite{gustafson2001advertising, crisp1987persuasive} (see Supplementary Note 2 for more details).

While it is essential to systematically examine racial and gender representation in visual media, it is equally important to understand the potential consequences of these portrayals on audience perceptions. Visual media are pervasive in daily life, and exposure to both inclusive (counter-stereotypical) and non-inclusive (stereotypical) imagery can shape how individuals perceive minority groups. Research has shown that stereotypical imagery can negatively influence viewers' attitudes and reinforce societal biases~\cite{ward2020media, reny2016negative, santoniccolo2023gender, casas2003impact, dong2007impact, behm2008mean}. For instance, exposure to situation comedies has been linked to skewed perceptions of African American education and income levels~\cite{tukachinsky2015documenting}, while stereotypical depictions of Native Americans have contributed to narrow identity prototypes, impacting self-perception within these communities~\cite{leavitt2015frozen}. In fashion, media representations have been found to affect self-perceptions of attractiveness and body image, advocating the need for more diverse and inclusive portrayals~\cite{makkar1995black, markova2018influence}. While much of the existing research has documented the negative effects of stereotypical imagery, fewer studies have explored the potential for counter-stereotypical portrayals to challenge ingrained biases~\cite{olsson2018does, savenye1990role}. Experiments that examine how exposure to inclusive versus non-inclusive media affects perceptions are particularly important, as they can provide insights into how repeated exposure to visual media can either reinforce or disrupt stereotypes.

Thus, to complement our observational analysis, we conduct three preregistered survey experiments with 1,800 participants to assess the impact of both inclusive (counter-stereotypical) and non-inclusive (stereotypical) media depictions on perceptions of racial and gender minorities. Given the pervasive nature of visual media, even short-term shifts in perception may contribute to longer-term changes in societal attitudes. These experimental results help us explore whether biased portrayals can reinforce or amplify societal stereotypes. If counter-stereotypical, inclusive content can reduce perception biases, even in the short term, the pervasive nature of media exposure could contribute to a longer-term reduction in societal bias, consistent with cultivation theory~\cite{morgan2014cultivation, nevzat2018reviving, lai2015media, shanahan1999television, romer2014cultivation}. By integrating large-scale observational data with experimental findings, our study not only provides a comprehensive overview of how racial and gender representations have evolved in different media formats but also offers an understanding of how visual media influence societal perceptions and biases, and how counter-stereotypical portrayals may offer a pathway toward greater inclusivity.

\section*{Results}
\subsection*{Vogue magazine}
\begin{figure}
    \centering
    \includegraphics{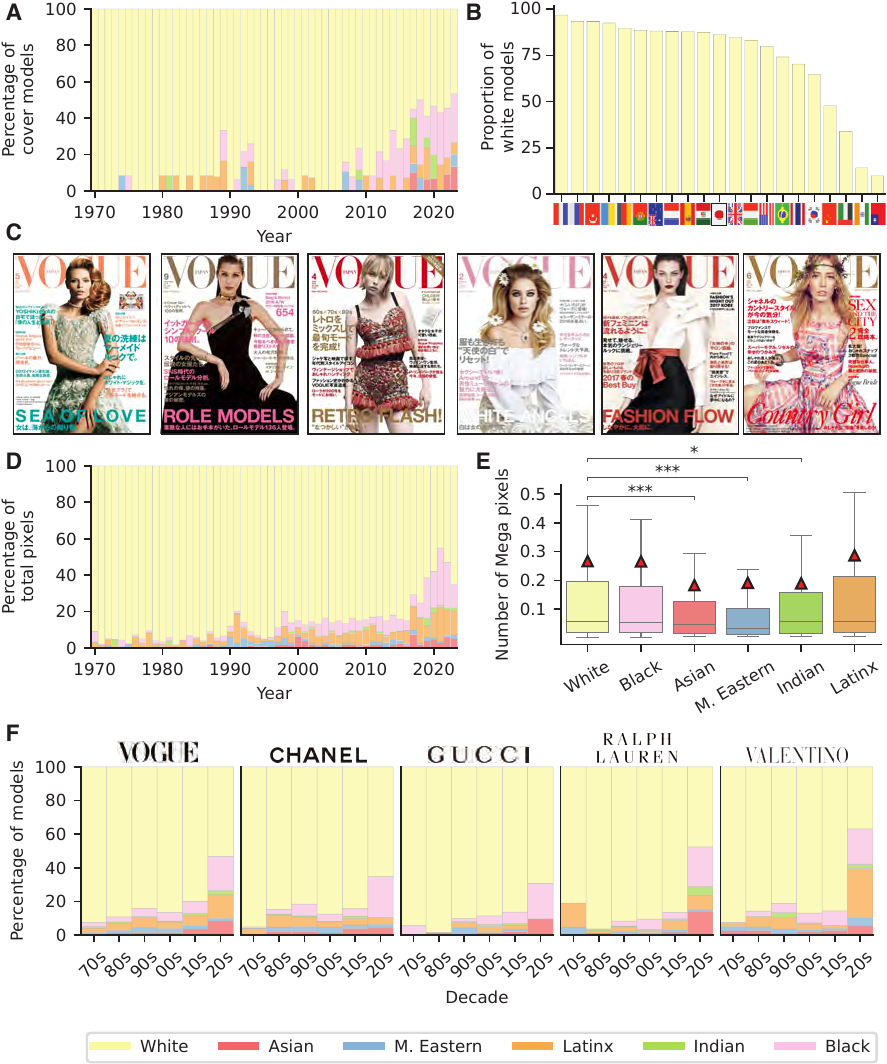}
    \caption{\textbf{Analysis of Vogue magazine photos.} (\textbf{A}) Distribution of cover model ethnicities over time in Vogue USA. (\textbf{B}) Proportion of White cover models in international Vogue magazines. (\textbf{C}) A sample of six Vogue Japan covers. (\textbf{D}) Distribution of number of pixels per year for model ethnicities in in-magazine photos of Vogue USA. (\textbf{E}) Box plots of the number of pixels allotted per image for each ethnicity among in-magazine photos of Vogue USA.  (Independent t-test:  (*: $p < 0.05$, **: $p < 0.01$, *** : $p < 0.001$). Comparisons made between White and all other racial groups, with only statistically significant comparisons shown. Means depicted with red triangle.) (\textbf{F}) Distribution of model ethnicities per decade among in-magazine photos in Vogue USA, as well as those published by four popular fashion houses.}
    \label{fig:fashion}
\end{figure}

We begin our exploration of racial diversity in visual media by analyzing the models featured across \textit{all} pages in \textit{Vogue} magazine, including both covers and in-page images (Figure~\ref{fig:fashion}; see Data Collection section for dataset details). First, we examine the racial distribution of models on the covers of \textit{Vogue USA} from 1970 to 2023 (Figure~\ref{fig:fashion}A). Over the decades, White models have overwhelmingly dominated the covers, with over 70\% of the covers featuring White models in all but one year, 1989, between 1970 and 2016. From 2016 to 2023, we observe a notable increase in non-White representation, especially for Black models, who appeared on more than 25\% of covers from 2019 to 2023. However, despite this progress, Asian, Middle Eastern, Latinx, and Indian models remain underrepresented, collectively accounting for only 20.8\% of the covers in the last five years.

To investigate whether these patterns extend internationally, we analyzed covers from 21 international editions of \textit{Vogue}. As shown in Figure~\ref{fig:fashion}B, White models dominate the covers globally, appearing on more than 80\% of the covers in 13 of the 21 editions and over 50\% in 17 of them. Surprisingly, this trend persists even in several non-White majority countries, such as South Korea (65\%), Thailand (70\%), and Japan (86\%). Figure~\ref{fig:fashion}C displays a sample of covers from \textit{Vogue Japan}, exemplifying how the dominance of White models is consistently reproduced, even in regions with predominantly non-White populations.

We next examine the racial diversity of models featured in \textit{in-page} photos of \textit{Vogue USA}. These images, unlike covers, vary in prominence, with models occupying different portions of the page. To quantify this, we measure the proportion of pixels occupied by each racial group, ensuring all pages were scanned at the same resolution. As shown in Figure~\ref{fig:fashion}D, White models consistently occupy the most space, similar to the observations on the covers (Figure~\ref{fig:fashion}A,B). Specifically, White models (\textit{N} = 34,590, \textit{M} = 0.26 megapixels, \textit{SD} = 0.59) occupy significantly more page area compared to Asian (\textit{N} = 881, \textit{M} = 0.18, \textit{SD} = 0.45), Middle Eastern (\textit{N} = 865, \textit{M} = 0.19, \textit{SD} = 0.48), and Indian models (\textit{N} = 369, \textit{M} = 0.19, \textit{SD} = 0.39). Figure~\ref{fig:fashion}E presents the direct comparisons of average space occupied by each racial group, calculated by the number of pixels allocated (White vs.\ Asian: \textit{D} $= 4.15$, $p < 0.001$; White vs.\ Middle Eastern: \textit{D} $= 3.71$, $p < 0.001$; White vs.\ Indian: \textit{D} $= 2.50$, $p = 0.012$).

Figure~\ref{fig:fashion}F further breaks down racial representation in in-page photos by advertisements from four major fashion houses: Chanel, Gucci, Ralph Lauren, and Valentino. Once again, White models are the most prominently featured across all four brands, even with the recent increase in minority representation overall. These patterns of racial disparity are consistently observed across all fashion houses featured in the \textit{Vogue USA} dataset (see Supplementary Table 1 for full details).

Overall, these findings reveal that White models have consistently been the most prominently featured group in \textit{Vogue} magazine, both in terms of frequency and prominence (i.e., page area occupied), across the USA and internationally over decades. Despite some improvements in the representation of non-White models, particularly Black models in recent years, significant racial disparities persist. This pattern extends beyond magazine covers to in-page imagery and fashion advertisements from leading brands. Given the influential role of major magazines like \textit{Vogue} in shaping cultural trends and beauty standards~\cite{reddy2018critical, millard2006stereotypes, markova2018influence}, the under representation of racial minorities could have broader implications for how beauty and fashion are perceived, especially in relation to racial minorities. In later sections, we further examine how these patterns in racial representation could impact public perceptions of beauty standards.

\subsection*{Hollywood movie casts}
Building on our analysis of racial diversity in \textit{Vogue} magazine, we now turn our attention to the diversity of casts in American films. We compiled a dataset of 12,907 movies produced by American production companies from 1960 to 2022, incorporating information from IMDb on cast members, including their billing order, date of birth, and photos (see Data Collection section for more details). Using this dataset, we impute the racial, gender, and age of the cast members and analyze how the diversity of these socio-demographic dimensions has changed over time in the context of American cinema, which holds a dominant influence on global film culture and has the most far-reaching impact on audiences worldwide~\cite{custard2003global, olson2000globalization, olson2001hollywood}.

\begin{figure}
    \centering
    \includegraphics[width = 0.9\linewidth]{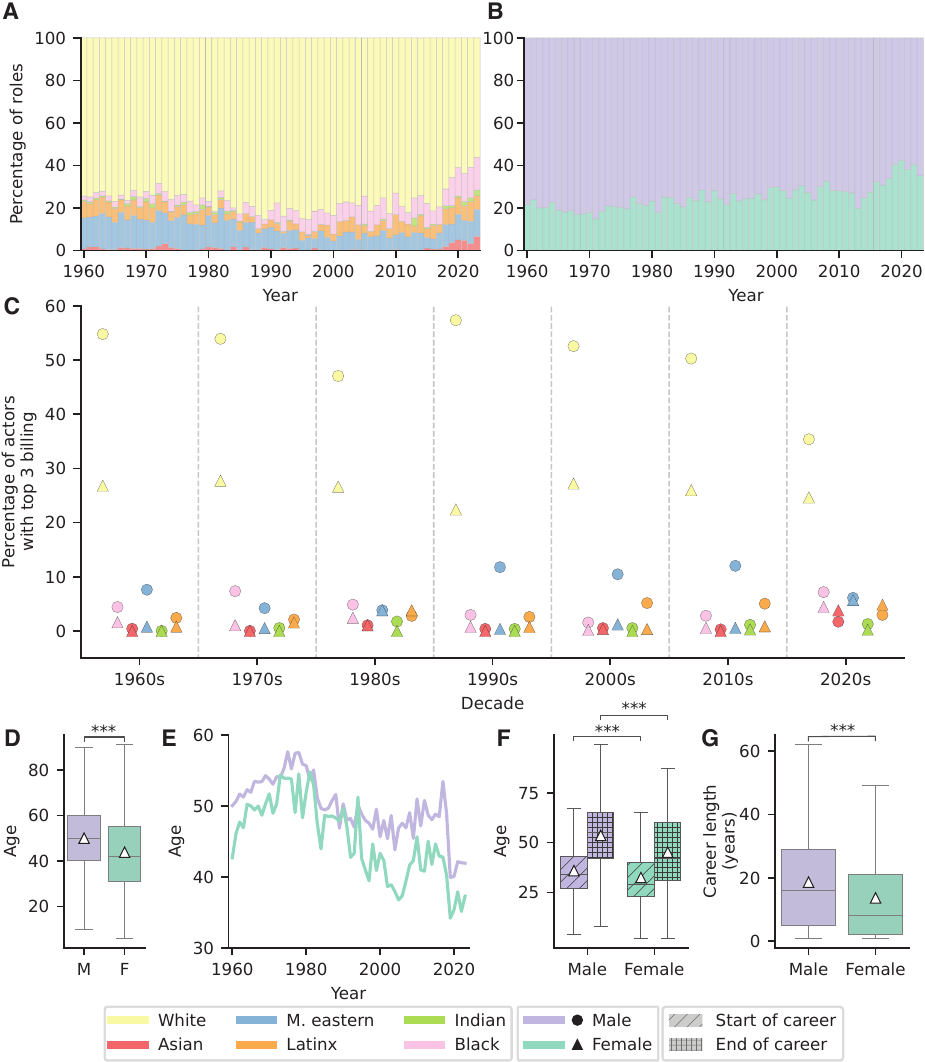}
    \caption{\textbf{Hollywood movie casts analysis.} (\textbf{A}) Percentage of roles filled by each race between 1960 and 2023. (\textbf{B}) Percentage of roles filled by female and male actors between 1960 and 2023.  (\textbf{C}) The percentage of all actors with a top three billing in a movie in a given decade of each race and gender. Females are depicted with a triangle, while males are depicted with a circle. (\textbf{D}) Box plots of the ages of female and male actors in movies. (\textbf{E}) The average age of female and male actors in movies over time.  (\textbf{F}) Box plots of the ages of female and male actors in movies at the start and end of their careers. (\textbf{G}) Box plots of the career lengths of female and male actors. For (\textbf{D}) , (\textbf{F}), and (\textbf{G}) independent t-tests are conducted (*: $p < 0.05$, **: $p < 0.01$, ***: $p < 0.001$).}
    \label{fig:movies}
\end{figure}

Figure~\ref{fig:movies}A shows the percentage of roles filled by actors from each racial group between 1960 and 2022. Note that the term `actors' refers to both female and male performers unless otherwise specified. White actors have historically dominated movie roles, peaking at 85.44\% of all roles in 1996, before decreasing to 56.19\% in 2023. Black actors have seen a steady increase in representation, reaching 15.43\% of all roles in 2023. Meanwhile, the representation of Asian actors remained low throughout the 20th century, only exceeding 3.5\% from 2019 onwards, and peaking at 6.65\% in 2023. We extend the same analysis to gender and observe that the gender breakdown of movie roles over time, where male actors have consistently held the majority of roles throughout the 63 years represented in our dataset (Figure~\ref{fig:movies}B). The highest representation of female actors occurred in 2019, when they held 40.4\% of all roles.

While the proportion of roles filled by racial and gender minorities provides a high-level view, it does not fully capture the importance of their roles within the context of the film. To address this, we analyze the share of top-billed roles (i.e., top 3 billing) occupied by cast members from each racial group (Figure~\ref{fig:movies}C). White male actors overwhelmingly dominate these top billing slots, followed by White female actors, whose top billing representation has remained relatively stable, fluctuating between 22\% and 29\%. Interestingly, minority groups exhibit different trends. The gap between female and male representation within these groups is more subtle compared to White actors, possibly due to the overall lower presence and prominence of minority actors in general. Moreover, for Latinx and Asian actors, female representation has surpassed that of their male counterparts in recent decades, a trend potentially linked to the increased hypersexualization of women from racial minority groups~\cite{taylor2018oversexualization, shimizu2007hypersexuality, tukachinsky2015documenting, glascock2018verbal}.

Next, we examine the intersection of gender and age in Hollywood to further explore how female and male cast members are portrayed differently. Figure~\ref{fig:movies}D highlights the average age of male and female actors at the time of their appearances in films, showing that male actors tend to be significantly older than their female counterparts (\textit{D} $ = 25.26, p < 0.001$). This age gap has even widened over time, as shown in Figure~\ref{fig:movies}E, with the difference in average age between male and female cast members increased from 3.9 years between 1960 and 1970 ($p < 0.001$) to 7 years between 2000 and 2010 ($p < 0.001$), despite both groups have tended to appear younger overall. This disparity can be driven by differences in both the starting and ending ages of acting careers. Surprisingly, male actors generally begin their acting careers later (35.9 years for males vs.\ 32.4 for females, $p < 0.001$), but continue acting until a significantly older age (53.6 years for males vs.\ 45.2 years for females, $p < 0.001$; Figure~\ref{fig:movies}F). Consequently, male actors tend to enjoy significantly longer careers, with an average career length of 18.7 years compared to 13.7 years for female cast members ($p < 0.001$), amounting to a 36\% longer career on average (Figure~\ref{fig:movies}G).

\subsection*{International movie posters}
While cast diversity in films provides insights into representation, the physical portrayals of characters on movie posters, such as body composition, posture, and facial expressions, offer additional layers of depicted prominence, power dynamics, and emotional states. Analyzing these visual cues complements our earlier findings by expanding our understanding of inequalities and discrepancies beyond cast listings to the visual hierarchy of characters on global marketing platforms. To explore these aspects, we collected posters from the 200 most popular movies per year released in 11 different languages, totaling 103,085 posters (see Methods for more details). Our goal is to investigate the portrayal of cast members in terms of body composition (which portion of their body is shown), posture (whether they are standing, sitting, or reclining), and the emotional expression conveyed through their facial features, which could potentially shape audience perceptions of characters' roles and significance.

\begin{figure}
    \centering
    \includegraphics[width = 0.90\linewidth]{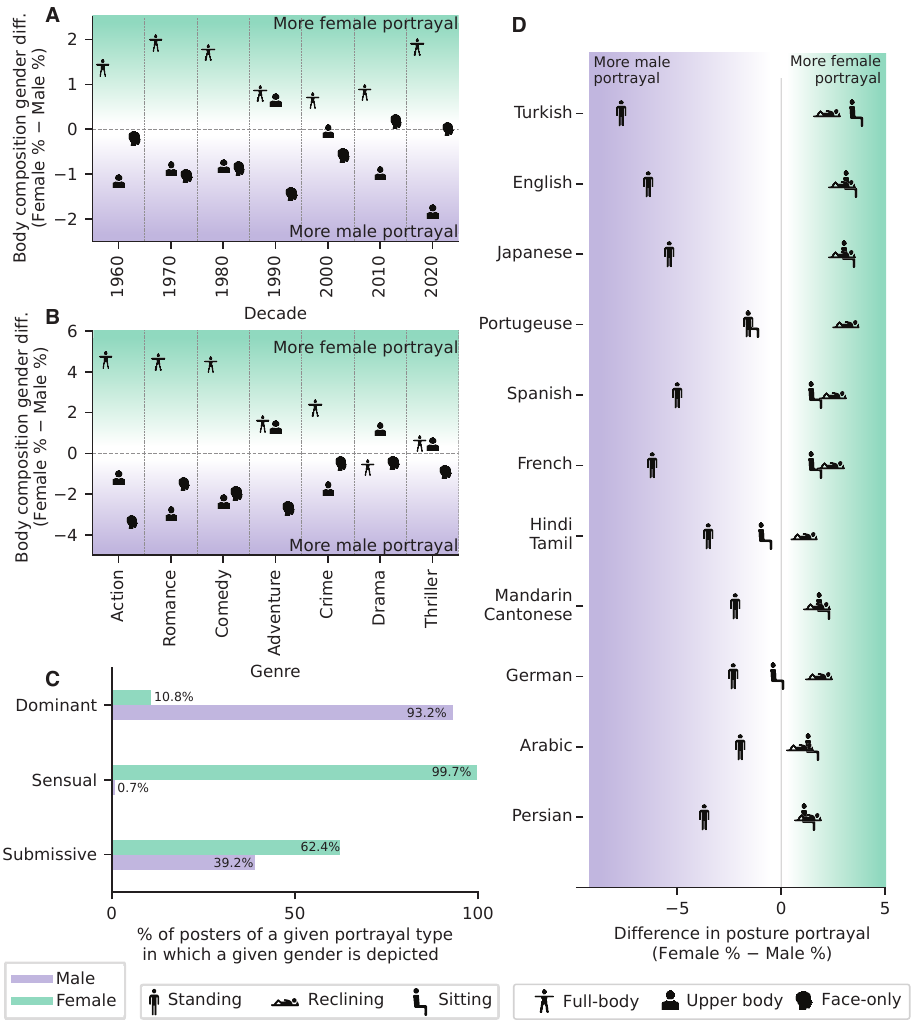}
    \caption{\textbf{Analysis of popular movie posters globally.} (\textbf{A}) The likelihood that a female actress is portrayed in a poster with a given portion of their body (face, upper body, and full body) minus the likelihood that a male actor is portrayed with that portion of their body. This analysis is grouped by decade for all global movie posters. (\textbf{B}) is similar to (\textbf{A}) but focusing on English movie posters and grouped by genre.  (\textbf{C}) The proportion of posters of a given portrayal type in which a given gender is depicted. (\textbf{D}) The likelihood that a female actress is portrayed in a poster with a given posture (standing, sitting, and reclining) minus the likelihood that a male actor is portrayed with that posture. This analysis is grouped by the primary language of the movie that a poster corresponds to.}
    \label{fig:posters}
\end{figure}

We begin by analyzing the body composition of cast members portrayed in these posters. Specifically, we measure whether male and female cast members are depicted in full-body, upper-body, or face-only shots (see Methods for more details on body composition analysis). These three categories dominate the posters, while other configurations are negligible (see Supplementary Table 2 for a detailed breakdown by language and Table 3 for body composition by decade). Figure~\ref{fig:posters}A shows that, across decades and languages, female actors are far more likely to be depicted with their full bodies visible in posters, while male actors are more frequently portrayed with only their face or upper body shown (Full body appearance: $\chi^2 = 15.03$, $p < 0.001$, Face appearance: $\chi^2 = 21.41$, $p < 0.001$, Upper body appearance: $\chi^2 = 6.07$, $p = 0.013$). This pattern holds across most genres (in movies where English is the primary language), except for dramas (Figure~\ref{fig:posters}B).

Next, we analyze the context in which male and female cast members are portrayed on these posters. Using a multi-modal large language model (GPT-4 Vision), we classify whether the cast members are depicted in dominant, sensual, or submissive poses (see Methods for more details on how these portrayal types are chosen and validated). As shown in Figure~\ref{fig:posters}C, female actors are overwhelmingly more likely to be portrayed in sensual or submissive postures (99.7\% sensual portrayals and 62.4\% submissive portrayals), while male actors are more frequently depicted in dominant postures (93.2\%). 

We further explore the postures of male and female cast members portrayed on these posters. Figure~\ref{fig:posters}D highlights the percentage difference between male and female cast members depicted in standing, sitting, and reclining positions. Males are more likely to be shown standing, particularly in Turkish, English, and French movie posters. In contrast, females are more frequently depicted in reclining positions, with this pattern persisting across all languages studied. Furthermore, female actors are more commonly portrayed in sitting positions in eight of the 11 languages analyzed. These results collectively suggest a clear gender divide in how authority, vulnerability, and emotional expression are visually conveyed on movie posters. Note that we replicate these results using an independent dataset of 14,184 advertisement images, further validating the robustness of our observations (see Supplementary Note 2).

As we emphasized earlier, both the diversity of representation and the physical portrayals can amplify or diminish how figures in cultural domains are perceived, often reinforcing cultural ideals of race, gender, and authority~\cite{browne1999television, clark2019seeing}. These portrayals may affect how audiences perceive power, status, and the importance of minority and gender groups, further entrenching or challenging existing biases~\cite{mastro2015media, tukachinsky2015we, browne1999television, ramasubramanian2011impact, shinoda2021beyond, ward2020media}. Given these dynamics, we conduct experiments to assess the effects of both stereotypical and counter-stereotypical images in a more controlled setting, aiming to uncover the extent of these biases and explore pathways to promote greater inclusivity through more balanced representation.

\subsection*{Survey experiments}
Given the pervasive presence of biased representations of racial and gender groups in popular culture and visual media, as demonstrated in Figures 1–3, we set out to examine how such imagery influences audience perceptions. Our objectives are twofold: (1) to determine whether individuals tend to rate the groups with greater exposure (i.e., White and male) more highly and (2) to assess whether exposure to more inclusive representations improves perceptions of minority groups. Among many possible experiments, we chose three specific settings that were clearly highlighted in our observational findings, each of which is outlined below. For each experiment, we compiled stimuli that closely mirror the stereotypical images observed in previous sections, which are widely encountered in everyday life. Additionally, we prepared counter-stereotypical stimuli featuring racially or gender-inclusive images to assess the impact of these alternative representations (see Methods for more details). These media formats are well-suited for experimental settings as they can be evaluated in real-time, simulating natural consumption experiences. Accordingly, we conducted three preregistered survey experiments to assess how exposure to inclusive (racially and gender-diverse) versus exclusive (White and male-dominated) media content influences participants' attitudes toward racial and gender groups, using fashion magazine covers and movie posters as stimuli. To further validate our findings, we run two additional sets of preregistered experiments focused on advertisements (see Supplementary Note 2).

Respondents (N = 600 for each experiment, totaling 1,800; mean age = 41.6 years;  52\% women) are recruited from the US via Prolific, an online platform providing a diverse pool of participants (see Methods section for more details). All survey materials are available in Supplementary Note 1, including participant demographics and response rates (Tables S10–14). 

\paragraph{Racial diversity in fashion magazine covers and perceived attractiveness}
The first experiment examines the effect of racially inclusive versus exclusive fashion magazine covers on participants' perceptions of attractiveness for different racial groups, particularly in relation to traditional beauty standards. This experiment is motivated by the persistent under representation of racial minorities in fashion magazines, as shown in Figure~\ref{fig:fashion}. While previous research has explored the relationship between media representation and self-perception, e.g., attractiveness and body dissatisfaction~\cite{makkar1995black, markova2018influence}, few studies have investigated how these images shape perceptions of the attractiveness of different racial groups. Given that fashion magazines often set societal beauty standards, we hypothesize that exposure to racially inclusive content will challenge conventional beauty norms and lead to an increased perceived attractiveness of minority groups.

In this experiment, participants are shown eight fashion magazine covers featuring either White models (exclusive condition) or a racially diverse set of models (inclusive condition), with distracting questions between the covers, or no images at all (baseline; see Methods section for more details). After exposure, participants rate the extent to which they associate the terms `classy,' `beautiful,' `attractive,' and `fashionable' with each racial group. These terms are chosen to assess both adherence to social beauty standards (`classy' and `fashionable') and general appeal (`beautiful' and `attractive').

\begin{figure}[hbt!]
    \centering
    \includegraphics[width = \linewidth]{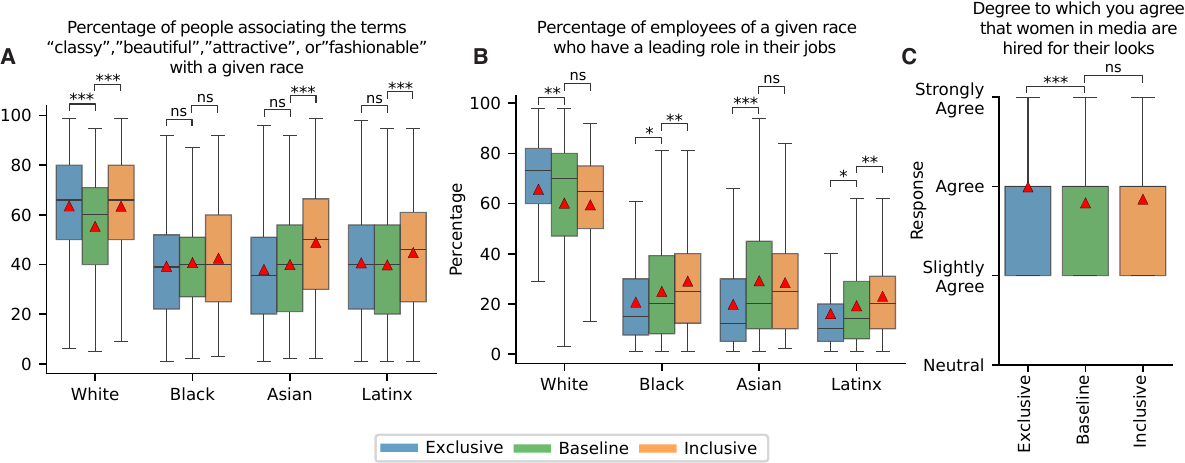}
    \caption{\textbf{Response distributions in survey experiments.} The three sub-figures correspond to the three survey experiments, and summarize participants' responses to the question listed above each sub-figure. In the baseline condition in all experiments, participants were not shown any images. (\textbf{A}) Response distributions when participants are shown fashion magazine covers with only white models (exclusive), diverse models (inclusive) or the baseline condition. (\textbf{B}) Response distributions when participants are shown movie posters with only white actors (exclusive), diverse actors (inclusive), or the baseline condition. (\textbf{C}) Response distributions when participants are shown movie posters with actresses portrayed in revealing attire (exclusive), actresses depicted in non-revealing attire (inclusive), or the baseline condition. (Mann Whitney U tests, ns: $p > 0.05$, *: $p < 0.05$, **: $p < 0.01$, ***: $p < 0.001$)}
    \label{fig:user_study}
\end{figure}

Figure~\ref{fig:user_study}A presents the results. When evaluating minorities, participants rate perceived attractiveness higher in the inclusive condition compared to both the baseline and exclusive conditions, though the differences for Black individuals are not statistically significant (Black: $D = 2.61$, $p = 0.930$; Asian: $D = 8.86$, $p < 0.001$; Latinx: $D = 5.14$, $p < 0.001$; all differences are in comparisons with the baseline). Interestingly, for all minorities, the exclusive and baseline conditions show no significant difference, potentially suggesting that the White dominance in the exclusive condition aligns with participants' pre-existing biases. When evaluating White individuals, participants in both the inclusive and exclusive conditions rate them more highly than in the baseline condition ($\mu =$ 64.14, 55.74, and 63.87, respectively, for exclusive, baseline, and inclusive conditions; $p < 0.001$ for all pairwise comparisons, except for the exclusive versus inclusive comparison, where $p = 0.660$).

We thus find supportive evidence that exposure to inclusive (counter-stereotypical) images increases the perceived attractiveness of minorities, particularly for Asian and Latinx groups. However, participants still rate the perceived attractiveness of White individuals as higher on average ($\mu = 61.31$) compared to other racial groups ($\mu = 41.64$, $42.45$, and $42.23$ for Black, Asian, and Latinx individuals, respectively; $p < 0.001$ for all pairwise comparisons), possibly reflecting prevalent preconceptions. 

\paragraph{Racial diversity in movie posters and perceived leadership}
In the second experiment, we explore how exposure to racially inclusive versus exclusive movie posters affects participants' perceptions of leadership roles associated with different racial groups. Prior research has shown that media portrayals can significantly affect self-identification with role models, particularly for individuals from underrepresented groups~\cite{hall2022audience}. This experiment aims to assess participants' estimates of how often racial minorities hold leadership positions in their respective domains. Given that media often reflects societal hierarchies~\cite{kendall2011framing}, we hypothesize that the portrayal of racial minorities in more prominent roles on movie posters would influence participants' beliefs about real-world leadership dynamics. Participants are shown eight movie posters either featuring only White actors (exclusive condition) or a racially diverse cast (inclusive condition), with distractor questions between images, or they view no images (baseline; see Methods section for more details).

As shown in Figure~\ref{fig:user_study}B, when evaluating minorities, participants estimate a higher percentage of Black and Latinx individuals in leadership roles after exposure to racially diverse movie posters (inclusive condition) compared to the baseline (Black: \textit{D} = 4.11, $p = 0.004$, Latinx: \textit{D} = 4.22, $p = 0.001$) while the effect was less clear when evaluating Asians (\textit{D} = -0.56, $p = 0.41$). Conversely, when exposed to White-only posters (exclusive condition), participants significantly underestimate the representation of minorities in leadership positions (Black: \textit{D} = 4.66, $p = 0.018$, Asian: \textit{D} = 9.53, $p < 0.001$, Latinx: \textit{D} = 2.94, $p = 0.025$). When evaluating White individuals, participants in the exclusive condition estimate an even higher proportion of White individuals in leadership positions compared to the baseline (\textit{D} = 6.00, $p = 0.004$), while exposure to diverse posters (inclusive condition) does not show a significant difference ($p = 0.101$).

Similar to Figure~\ref{fig:user_study}A, we find supportive evidence that exposure to inclusive (counter-stereotypical) images increases the perceived presence of minorities in leadership roles, particularly when compared to perceptions after exposure to exclusive (stereotypical) images. However, participants continue to estimate that White individuals hold more leadership positions on average ($\mu = 62.18$) compared to other racial groups ($\mu = 24.77$, $25.65$, and $19.17$ for Black, Asian, and Latinx individuals, respectively; $p < 0.001$ for all pairwise comparisons). 

\paragraph{Female attire in movie posters and perceived appearance-based casting}
The third experiment examines how the portrayal of female actors in movie posters---either in revealing or non-revealing attire---affects participants' views on appearance-based casting in media. Prior research has highlighted the hypersexualization of women in media~\cite{tukachinsky2015documenting, glascock2018verbal} and has shown that exposure to sexualized images of women, such as in superhero films, can lead to lower body esteem~\cite{pennell2015empowering}. This experiment assesses whether such sexualized portrayals reinforce the stereotype that female actors are cast based primarily on their looks. Grounded in prior research that demonstrates how sexualized depictions of women perpetuate the notion that their value is tied to appearance~\cite{bernard2012integrating,ward2016media,ward2023sources}, we hypothesize that exposure to movie posters featuring female actors in revealing attire will activate appearance-based stereotypes, reinforcing the belief that women in media are cast for their looks, whereas non-revealing attire will not trigger these biases.

Participants are shown eight movie posters featuring female actors in either revealing attire (exclusive condition) or non-revealing attire (inclusive condition). They are then asked to rate their agreement with the statement: ``Women in media are cast for their looks'' on a 7-point Likert scale. As shown in Figure~\ref{fig:user_study}C, participants exposed to the exclusive condition (revealing attire) are significantly more likely to agree with the statement compared to those in the baseline condition (\textit{D} = 0.34, $p < 0.001$). In contrast, no significant difference is observed between the inclusive (non-revealing attire) condition and the baseline ($p = 0.460$). These findings suggest that portrayals of female actors in revealing attire reinforce the stereotype that women in media are cast based on appearance, while non-revealing portrayals do not contribute to this stereotype.

\section*{Discussion}
This study examines racial and gender biases across visual media at an unprecedented scale, focusing on fashion magazines, movie casts, and posters, with additional analyses on advertisements provided in the Supplementary Information. Our analysis not only assesses the frequency with which different racial and gender groups appear in media but also investigates how these groups are portrayed. In terms of racial representation, we find that White individuals overwhelmingly dominate appearances across all media forms, both in the United States and globally, as seen in international issues of \textit{Vogue} magazine. Moreover, when racial minorities are featured, their portrayal often differs: they are depicted in smaller areas within fashion magazines, given less important roles in films, or convey less positive emotions in advertisements. For gender, our findings consistently show that women are more likely to be depicted in full-body images and in sensual or submissive poses. These depictions are concerning, given prior research showing that individuals presented in full-body images are often perceived as less intelligent and less likable compared to those shown only from the shoulders up, a concept termed by the literature as \textit{face-ism}~\cite{schwarz1989s, gray2011more,  konrath2012cultural}. Such biases are particularly troubling in visual media or in images retrieved from search engines~\cite{ulloa2024representativeness}, where recent work suggests that gender biases may be more potent in images than in text~\cite{guilbeault2024online}. These patterns may also explain the gender biases observed in AI-generated images~\cite{sun2024smiling}, which have the potential to broadly reproduce and amplify societal biases.

Our experimental results demonstrate that media can significantly influence public perceptions of racial and gender groups. Exposure to inclusive content often reduces biases, while stereotypical White-dominant and female-sexualized portrayals reinforce them. Specifically, the experiments reveal that racially diverse fashion magazine covers can increase perceived attractiveness of minority groups, particularly Asian and Latinx individuals. Conversely, the persistent dominance of White individuals in fashion media aligns with participants’ preconceptions about beauty standards, underscoring the role of media in shaping societal norms. In the context of movie posters, the representation of racial minorities in movie posters influences participants' perceptions of leadership roles. Viewing racially diverse posters led participants to estimate higher proportions of minorities in leadership positions, suggesting that diverse media can shift perceptions of authority and representation in broader societal contexts. Moreover, our findings indicate that portrayals of women in revealing attire reinforce stereotypes that women in media are valued primarily for their appearance. However, non-revealing portrayals do not contribute to this bias, suggesting that more neutral depictions can challenge entrenched stereotypes. This aligns with previous research demonstrating the positive impact of counter-stereotypical imagery on gender role beliefs~\cite{pennell2015empowering, finnegan2015counter}.

While our study focuses on short-term shifts in perception, rather than the long-term effects of exposure to inclusive or exclusive content, the findings align with cultivation theory~\cite{morgan2014cultivation, nevzat2018reviving, lai2015media, shanahan1999television, romer2014cultivation}, which suggests that repeated exposure to media signals can shape societal values over time. This implies that inclusive, counter-stereotypical content has the potential to promote equity and reduce biases in the long run. Given that visual mediums, such as fashion magazines, movie posters, and advertisements, remain strongly biased, our study highlights both the potential benefits of diverse, counter-stereotypical portrayals of racial and gender minorities and the risks posed by content dominated by specific racial or gender groups.

It is important to note that our study focuses on perceived gender and race, rather than self-identified data, to better understand how media representations are shaped by perceived identities and how these perceptions influence biases. The models used in our study classify race into seven categories (Black, East Asian, Southeast Asian, Indian, Latinx, Middle Eastern, and White) and gender as binary (male or female) based on the available labels in the training data. This typical limitation restricts our ability to account for other racial and gender identities. Future research could address this limitation by incorporating more advanced classifiers that reliably capture a broader range of racial and gender groups, allowing for more nuanced analyses of representation and bias across various media on a larger scale.

Additionally, our datasets, with few exceptions, primarily focus on Western media, which may limit the applicability of our findings in other cultural contexts. Future studies could explore similar biases in different regional contexts to provide a more holistic understanding of global representation. While our analysis centers on print media and film, exploring other media formats---such as online news, social media, or television---could reveal further insights into the prevalence and impact of bias across different platforms.

Finally, our experiments serve as a broad survey of general tendencies rather than isolating specific psychological mechanisms. This exploratory approach is still valuable for identifying bias patterns following media exposure and offers a foundation for more targeted, mechanism-focused studies. In real-world applications, such as media studies, consumer research, or public policy, understanding general audience reactions is critical for informing large-scale decisions. By mapping these broad responses, our work contributes to theory-building and sets the stage for future research that investigates causal processes and specific mechanisms more deeply.

\section*{Methods}
\subsection*{Data collection}
\subsubsection*{Fashion magazines}
We collected data from \textit{Vogue USA} by purchasing the full archive of 648 issues spanning the years 1970 to 2023 via ProQuest. This included both the covers and all internal pages, with detailed annotations identifying the specific fashion houses featured on each page. In total, our dataset consists of 751 models featured on the covers and 41,222 models depicted within the pages of these issues. For the international editions of \textit{Vogue}, cover data was sourced from the Vogue Covers archive~\cite{vogue_archive}, encompassing 4,064 models across 4,059 issues from 21 different international editions.

\subsubsection*{Movie casts}
To compile data on Hollywood movie casts, we began by retrieving yearly lists of movies released by American film studios between 1960 and 2023 from Wikipedia~\cite{wiki_2024}. Cast information for each movie was then collected from the IMDb database. For each unique cast member, we extracted their photograph and date of birth from their IMDb profile. In cases where this information was unavailable on IMDb, we supplemented it using the bio section of the actor's Wikipedia page. Actors lacking this information on both platforms were excluded from our analysis. To determine each actor's gender, we searched for gender-identifying pronouns on their IMDb and Wikipedia pages. Where this data was missing, we used the Genderize Python package to infer gender based on their name~\cite{genderize_2019}. This process yielded a dataset of 12,907 movies and 7,086 cast members.

\subsubsection*{Movie posters}
For global movie posters, we queried the TMDB API~\cite{tmdb} to retrieve the 200 most popular movies per year, released in one of eleven languages: Arabic, English, French, German, Hindi, Japanese, Mandarin, Persian, Portuguese, Spanish, and Turkish. We focused on these popular films because their posters are likely to have the greatest visibility and influence on audiences. High-quality poster images were collected through URLs provided by TMDB for films released between 1950 and 2023, yielding a total of 103,085 posters. Since our analysis centers on racial and gender representation, we excluded 2,646 posters that did not feature any humans with detectable demographic attributes, resulting in a final dataset of 100,439 posters.

\subsection*{Image classifiers}
\subsubsection*{Race, gender, and emotion classification models}
Recent advancements in automated demographic inference methods have been applied in a variety of contexts~\cite{amaral2020long,lockhart2023name,park2023fighting,aldahoul2024ai}. Despite these innovations, such approaches remain underutilized in media representation research, with some notable exceptions~\cite{amaral2020long}. Our study leverages these developments by applying state-of-the-art machine learning techniques to extract nuanced demographic information from images. This enables us to classify not only race and gender but also the posture, body composition, and emotional expression of individuals depicted in visual media.

To classify race and gender, we use the model recently proposed by Al Dahoul et al.~\cite{aldahoul2024ai}, which is trained on the FairFace dataset~\cite{karkkainenfairface}, one of the largest publicly available face image datasets. Importantly, FairFace was specifically designed to be balanced across racial and gender groups, addressing potential biases arising from imbalanced training data. The model classifies faces into one of seven racial groups: Black, East Asian, Indian, Latinx, Middle Eastern, Southeast Asian, and White. These categories follow those chosen by FairFace, constructed on classifications commonly used by the U.S. Census Bureau (White, Black, Asian, Hawaiian and Pacific Islanders, Native Americans, and Latino), with further subdivisions for groups like Middle Eastern, East Asian, Southeast Asian, and Indian to capture distinct appearances. Due to the limited number of examples for Hawaiian and Pacific Islanders and Native Americans in the dataset, these categories were excluded, resulting in the seven racial groups used in our analysis. For simplicity, we group East Asian and Southeast Asian into a single category, labeled `Asian.' 

As demonstrated in the original research~\cite{aldahoul2024ai}, this model outperforms other popular classifiers, including Google's FaceNet~\cite{schroff2015facenet} and CLIP's zero-shot classifier~\cite{radford2021learning}. We also use this model to classify gender when explicit information is unavailable, such as in cases where models are depicted in advertisements.

To classify the emotional state conveyed by a subject in any given image, we use the POSTER++ model developed by Mao et al.~\cite{mao2023poster++}, which categorizes facial expressions into one of eight emotional states: neutral, happy, sad, surprise, fear, disgust, anger, and contempt. In this model, images are first cropped to isolate the subject’s face before classification. While facial recognition models like this have been shown to potentially exhibit racial biases ~\cite{wolfe2022evidence}, we find the performance reported by the authors to be sufficient for the purposes of our study.

\subsubsection*{Portrayal, posture, and body composition identification}
To gain deeper insight into the ways in which individuals are portrayed in images, we use GPT-4 with Vision~\cite{openai_2024}. Unlike models above that crop images to isolate subjects, we prompt GPT-4 on uncropped images using the \textit{GPT-4 Vision Preview} model, with a temperature set to 0.5. This intermediate temperature value was chosen as studies show that temperature does not significantly impact performance for problem-solving tasks in large language models~\cite{renze2024effect}. The following prompt was used to guide the model:

\begin{quote}
    \textit{Answer each of the following questions in no more than 7 words:}
    \begin{enumerate}
        \item \textit{Is there a person shown in the image? Answer with one of [`yes', `no'].}
        \item \textit{What is the gender and posture of each person in the image? Separate each pair of gender and posture with a `\&' sign.}
        \item \textit{For each person, show the gender and select one option only from [`full body', `upper body without legs', `lower body only', `face only', `body without face']. Separate each pair of gender and option with a `\&' sign.}
        \item \textit{How are the female(s) being portrayed?}
        \item \textit{How are the male(s) being portrayed?}
    \end{enumerate}
\end{quote}

To further investigate how different gender groups are portrayed, we analyzed the proportion of images depicting each gender in specific portrayal contexts. First, we compiled GPT-4's responses to Questions 4 and 5 from the prompt above for all images featuring subjects of a given gender. We then computed vectorized embeddings of these descriptions using OpenAI's \textit{text-embedding-3-large} model. With these embeddings, we applied k-means clustering to group similar descriptions in the embedding space. Specifically, we set $k = 20$ to capture a broad range of portrayal types for subjects within an image. Following this step, to identify overarching labels for each of the 20 clusters, we randomly sampled 30 descriptions from each cluster and fed them into GPT-4 with the prompt: \textit{``Summarize the following descriptions using one to two words.''} This process yielded the following cluster labels: neutral, sensual, submissive, dominant, distressed, casual, serious, confident, happy, sad, angry, playful, fearful, disgusted, surprised, positive, negative, active, nurturing, and calm.

Next, one of the first authors manually inspected the images in each of the 20 clusters to confirm whether the subjects were consistently portrayed according to the associated cluster labels. From this process, three clusters of particular interest emerged as the most coherent: portrayals that were sensual/sexual, submissive, or dominant. Using these categories as our targets, we identified relevant images for each by compiling a set of 10 synonyms for each cluster label. These synonyms were selected by finding the 10 most similar adjectives and adverbs to each label in the vector embedding space using FastText's English embedding model~\cite{fast_text}, resulting in 20 keywords per cluster. We then filtered images based on whether GPT-4's descriptions contained any of these keywords, isolating those that aligned with one of the three portrayal clusters. An image could be associated with multiple clusters if it contained keywords from more than one. The keywords for each cluster were as follows: (i) \textit{Sensual/Sexual}: sensuously, sensuous, erotically, erotic, seductively, seductive, sexily, sexy, sexually, sexual, voluptuously, voluptuous, lasciviously, lascivious, alluringly, alluring, salaciously, salacious, hypnotically, hypnotic; (ii) \textit{Submissive}: submissive, submissively, subserviently, subservient, obediently, obedient, docilely, docile, meekly, meek, obsequiously, obsequious, demurely, demure, supinely, supine, timorously, timorous; and (iii) \textit{Dominant}: dominantly, dominant, domineering, domineeringly, commandingly, commanding, imperiously, imperious, imposingly, imposing, measuredly, measured, demandingly, demanding, overbearingly, overbearing, bossily, bossy, tyrannically, tyrannical.

The output quality of these models---for identifying posture, body part shown, and portrayal type---was verified by three independent coders. As shown in Table~\ref{table:annotation}, GPT-4 successfully identified these attributes with an accuracy of at least 81\%. Inter-rater agreement was assessed using Krippendorf's $\alpha$, with all annotations achieving $\alpha > 0.86$, indicating strong consensus among coders. Additionally, Cohen's $\kappa$ was calculated between the majority label of the independent coders and GPT-4's outputs, with each cluster achieving $\kappa > 0.6$, further supporting the reliability of the model's classifications.

\subsection*{Survey experiment procedure}
To measure the effects of exposure to inclusive versus non-inclusive media images on perceptions of prominence, attractiveness, and other attributes related to each racial group, we conducted three preregistered experiments. These experiments are detailed in the results section, with additional analyses provided in Supplementary Note 2. The first two experiments aimed to assess the impact of racially inclusive versus White-dominant media images compared to a baseline condition where participants were not shown any images. These experiments used two distinct media forms: movie posters and fashion magazine covers. Participants were shown a set of eight images, interspersed with distractor questions (see Supplementary Tables 15-24 for responses to distractor questions), and then asked a series of questions to gauge their perceptions regarding the specific outcome of interest. The third experiment examined how exposure to movie posters depicting female actors in revealing attire versus non-revealing attire affected participants' views on the role of appearance in casting decisions for female actors.

For the first two experiments, participants in the exclusive condition viewed eight \textit{Vogue USA} covers randomly sampled from those featuring White models. In the inclusive condition, two covers were randomly selected for each racial group from the set of \textit{Vogue USA} covers featuring models from four racial backgrounds used in our analysis: White, Black, Asian, and Latinx. In the third experiment, movie posters in the exclusive condition were sampled from those labeled as portraying female actors in a sexualized manner (see Figure~\ref{fig:posters}C), while those in the inclusive condition were sampled from posters featuring female actors in non-revealing attire.

To achieve a statistical power of 0.8 for detecting a moderate effect size (Cohen's d = 0.25), we estimated that 199 participants per condition would be required. Therefore, 200 participants were recruited for each of the exclusive, inclusive, and baseline conditions in all studies. Participants were sourced from the United States using the Prolific platform~\cite{prolific_2014}.

The full set of survey materials can be found in Supplementary Note 1. The images shown to participants in each of the five studies are presented in Supplementary Figures 4-13. Participant demographics are detailed in Supplementary Tables 10–14, and response rates for each survey question are provided in Supplementary Tables 5–9. The preregistration information for these experiments is available at: \url{https://aspredicted.org/a87yx.pdf}.

\section*{Ethics Statement}
The research was approved by the Institutional Review Board of New York University Abu Dhabi (HRPP-2024-57). All research was performed in accordance with relevant guidelines and regulations. Informed consent was obtained from all participants in every segment of this study.


\clearpage

\clearpage

\begin{table}[h!]
\centering
\renewcommand{\arraystretch}{1.4} 
\begin{tabular}{l|>{\centering\arraybackslash}m{0.15\textwidth}|>{\centering\arraybackslash}m{0.15\textwidth}|>{\centering\arraybackslash}m{0.35\textwidth}}
\hline
~ & \textbf{Posture} & \textbf{Body Parts} & \textbf{Sensual, Submissive, or Dominant Portrayal} \\ \hline
\textbf{Accuracy (\%)} & 93.5 & 81.1 & 88.1 \\
\textbf{F1 Score} & 0.95 & 0.81 & 0.88 \\
\textbf{Rater-AI agreement} & $0.64^{***}$ & $0.85^{***}$ & $0.82^{***}$ \\
\textbf{Inter-rater agreement} & $0.901^{***}$ & $0.904^{***}$ & $0.863^{***}$ \\ \hline          
\end{tabular}
\caption{Model accuracy, F1 score, rater-AI agreement (Cohen’s $\kappa$), and inter-rater agreement (Krippendorff’s $\alpha$) for detected body parts and portrayed postures and portrayals (sensual, submissive, or dominant), compared with human coders using GPT-4's classification. (*: $p < 0.05$, **: $p < 0.01$, ***: $p < 0.001$)}
\label{table:annotation}
\end{table}

\newpage

\bibliographystyle{naturemag}
\bibliography{sample}

\newpage

\section*{Data and code availability}
All machine learning model outputs and code to generate the figures can be found at ~\url{github.com/comnetsad/bias_in_media}. Raw image data can be obtained from the corresponding author with reasonable request.

\section*{Author Contributions}
T.R. and Y.Z. conceived the study and designed the research; N.A., H.I., and M.P. collected the data; N.A. developed the machine learning models and applied them on the data; N.A. and H.I. analyzed the data; H.I., M.P., T.R., and Y.Z. wrote the manuscript.

\section*{Competing Interests}
The authors declare no competing interests.

\end{document}